\documentstyle[aps,prd,preprint]{revtex}
\begin{document}
\hfill{NCKU-HEP-98-09}\par
\hfill{hep-ph/9811340}
\vskip 0.3cm
\begin{center}
{\large {\bf $k_T$ and threshold resummations}}
\vskip 1.0cm
Hsiang-nan Li
\vskip 0.3cm
Department of Physics, National Cheng-Kung University, \par
Tainan 701, Taiwan, Republic of China
\end{center}
\vskip 1.0cm

\vskip 1.0cm

\centerline{\bf Abstract}
\vskip 0.3cm
We demonstrate that both the $k_T$ and threshold resummations can be
performed in the Collins-Soper resummation formalism by evaluating soft
gluon emissions with infrared cutoffs for the longitudinal and
transverse loop momenta, respectively. The reason the $k_T$ resummation for
a parton distribution function leads to suppression in the large $b$
region, $b$ being the conjugate variable of parton transverse momentum
$k_T$, and the threshold resummation leads to enhancement in the large
$N$ limit, $N$ being the moment of a distribution function, is a
consequence of opposite directions of double-logarithm evolutions.
The $k_T$ and threshold resummations for an energetic final-state jet give
suppression. The switch of the threshold resummation from enhancement to
suppression is attributed to a nonvanishing jet invariant mass. In the
same framework we derive a unification of the $k_T$ and threshold
resummations for a parton distribution function by requiring infrared
cutoffs for both longitudinal and transverse loop momenta. This unified
resummation exhibits suppression at large $b$, similar to the $k_T$
resummation, and exhibits enhancement at small $b$, similar to the threshold
resummation.

\newpage
\centerline{\large \bf I. INTRODUCTION}
\vskip 0.5cm

Recently, we have applied the Collins-Soper (CS) resummation technique
\cite{CS} to the $k_T$ resummation for a parton distribution function
$\phi(\xi,k_T,p^+)$ associated with a hadron of momentum
$p^\mu=p^+\delta^{\mu+}$ \cite{L1}, which describes the probability that
a parton carries longitudinal momentum $\xi p^+$ and transverse
momentum $k_T$. In this formalism the double logarithms
$\ln^2(p^+b)$ contained in $\phi$, $b$ being the conjugate variable of
$k_T$, are organized. The result is a Sudakov suppression factor,
quoted as \cite{L1}
\begin{equation}
\phi(\xi,b,p^+)=\exp\left[-\int_{1/b}^{\xi p^+}\frac{d p}{p}
\int_{1/b^2}^{p^2}\frac{d\mu^2}{\mu^2}
\gamma_{K}(\alpha_s(\mu))\right]\phi(\xi)\;,
\label{fb0}
\end{equation}
where the anomalous dimension $\gamma_K$ will be defined in Sec. II, and
the initial condition $\phi(\xi)$ of the double-logarithm evolution
coincides with the standard parton model. We have neglected the dependence
of $\phi$ on the renormalization (or factorization) scale $\mu$, which is
not relevant to our discussion here. The $\mu$ dependence denotes a
single-logarithm evolution, which can be easily derived using
renormalization-group (RG) equations.

Equation (\ref{fb0}) has been employed to evaluate the $p_T$ spectrum of
direct photon production, $p_T$ being the transverse momentum of the direct
photon \cite{LL}. It was observed that the double-logarithm exponential,
after Fourier transformed to $k_T$ space, provides the necessary
$k_T$ smearing effect, which resolves the discrepancy between most data sets
and next-to-leading-order QCD ($\alpha\alpha_s^2$) predictions
\cite{HKK}. This success implies the importance of the $k_T$ resummation for
studies of hard semi-inclusive QCD processes, in which transverse
degrees of freedom of final states are measured.

In a series of works \cite{S} the threshold resummation \cite{S0,CT,KM} for
dijet, direct photon and heavy quark productions in kinematic end-point
regions has been performed. In this formalism a different type of double
logarithms $\ln^2(1/N)$, which come from the Mellin transformation of the
logarithmic corrections $\ln(1-\xi)/(1-\xi)_+$ to moment $N$ space, is
organized. They are important logarithms, since $\xi\to 1$
corresponds to the large $N$ limit. On the contrary, the threshold
resummation for a parton distribution function $\phi(\xi,p^+)$, where the
transverse degrees of freedom have been integrated out, results in
enhancement:
\begin{eqnarray}
\phi(N,p^+)=\exp\left[-\int_{0}^{1}dz\frac{z^{N-1}-1}{1-z}
\int_{(1-z)^2}^{1}\frac{d\lambda}{\lambda}
\gamma_{K}(\alpha_s(\sqrt{\lambda} p^+))\right]\phi^{(0)}\;,
\label{ft2}
\end{eqnarray}
with the initial condition $\phi^{(0)}$ and
the same anomalous dimension $\gamma_K$.

At the kinematic end points of QCD processes, energetic final-state jets may
be produced, which also involve large double logarithms. Take deep inelastic
scattering (DIS) as an example. The final-state quark scattered by the
photon with momentum $q$ carries the momentum
$p'=\xi p+q=((\xi-x)p^+,q^-,{\bf 0})$ for the
Bjorken variable $x=-q^2/(2p\cdot q)=-q^+/p^+$ in a frame where $q$
possesses only the plus and minus components. As $x\to 1$, $p'$ approaches
the light cone with vanishing invariant $p^{\prime 2}$, and the
final-state quark forms a jet. The threshold resummation for a jet function
$J(N,p^{\prime-})$ gives Sudakov suppression \cite{S},
\begin{eqnarray}
J(N,p^{\prime-})=\exp\left[\int_{0}^{1}dz\frac{z^{N-1}-1}{1-z}
\int_{(1-z)^2}^{1-z}\frac{d\lambda}{\lambda}
\gamma_{K}(\alpha_s(\sqrt{\lambda} p^{\prime-}))\right]J^{(0)}\;,
\label{jt2}
\end{eqnarray}
opposite to the enhancement in Eq.~(\ref{ft2}). While the $k_T$ resummation
for a jet function $J(b,p^{\prime-})$ still gives Sudakov suppression
similar to Eq.~(\ref{fb0}) \cite{LY},
\begin{equation}
J(b,p^{\prime-})=\exp\left[-\int_{1/b}^{p^{\prime-}}\frac{d p}{p}
\int_{1/b^2}^{p^2}\frac{d\mu^2}{\mu^2}
\gamma_{K}(\alpha_s(\mu))\right]J^{(0)}\;.
\label{fj0}
\end{equation}
The above expression has been intensively appied to the analyses of
end-point spectra of heavy meson decays \cite{LY}.

With the above comparision, it is worthwhile to explore the relation
between the $k_T$ and threshold resummations, and the difference
between their results for a parton distribution function and for a jet
function. In this paper we shall show that the threshold resummation can
also be performed in the CS framework. The double
logarithms $\ln^2(p^+b)$ and $\ln^2(1/N)$ are summed simply by choosing
appropriate infrared cutoffs in the evaluation of soft gluon corrections.
If transverse degrees of freedom of a parton are included, $1/b$ will serve
as a transverse infrared cutoff. Combined with the
collinear logarithms $\ln p^+$, $\ln^2(p^+b)$
are generated. In the end-point region with $x\to 1$, $p^+/N$ is a
longitudinal infrared cutoff, implying the presence of
$\ln^2(1/N)$. Therefore, in the region away from end points, we neglect
the longitudinal cutoff, and sum $\ln^2(p^+b)$. As close to end points, we
keep only the longitudinal cutoff, and integrate over transverse
degrees of freedom of a parton in the same formalism. The logarithms
$\ln^2(1/N)$ are then summed.

It will be shown, in the CS framework, that suppression from the $k_T$
resummation and enhancement from the threshold resummation for a
parton distribution function is a consequence of opposite directions
of double-logarithm evolutions: from the small $1/b$ to the large $p^+$
in the former case and from $p^+$ to $p^+/N\ll p^+$ in the latter case. The
comparision of the two resummations for a jet function is subtler.
In the $k_T$ resummation the jet momentum is parametrized as
$p'=(0,p^{\prime-},{\bf k}_T)$, similar to the parton momentum
$(\xi p^+,0,{\bf k}_T)$, and thus the same suppression effect is expected. 
In the threshold resummation the jet momentum is 
$p'=((\xi-x)p^+,q^-,{\bf 0})$, different from the parton momentum
$(\xi p^+,0,{\bf 0})$. The former has a nonvanishing invariant mass
$p'^2=2(\xi-x)p\cdot p'$, while the latter lies on the light cone. We shall
demonstrate that this difference is responsible for the
switch from enhancement for a parton distribution function to
suppression for a jet fucntion.

Once the two different resummations can be reproduced in the same formalism,
it is possible to develop a unified resummation, in which both the
logarithms $\ln^2(p^+b)$ and $\ln^2(1/N)$ are organized to all orders. The
unification of the $k_T$ and threshold resummations for a parton
distribution function are achieved simply by retaining the longitudinal and
transverse cutoffs simultaneously. It will be observed that the result
exhibits suppression at large $b$, similar to the $k_T$ resummation, and
enhancement at small $b$, similar to the threshold resummation. The unified
resummation is appropriate for studies of QCD processes, in which
final states possess large rapidities (corresponding to large momentum
fractions), and their transverse degrees of freedom are measured.

In Sec. II we review the application of the CS technique to the $k_T$
resummation for a quark distribution function. In Sec. III we perform the
threshold resummation for a quark distribution function in the
same formalism, comparing each step with that in the $k_T$ resummation.
The threshold resummation for a jet function is derived in Sec. IV.
The unification of the two resummations is proposed in Sec. V.
Section VI is the conclusion.

\vskip 1.0cm

\centerline{\large \bf II. $k_T$ RESUMMATION}
\vskip 0.5cm

Consider a quark distribution function for a hadron in the minimal
subtraction scheme,
\begin{equation}
\phi(x,k_T,p^+)=\int\frac{dy^-}{2\pi}\int\frac{d^2y_T}{(2\pi)^2}
e^{-ix p^+y^-+i{\bf k}_T\cdot {\bf y}_T}
\langle p| {\bar q}(y^-,{\bf y}_T)\frac{1}{2}\gamma^+q(0)|p\rangle\;,
\label{dep}
\end{equation}
where $\gamma^+$ is a Dirac matrix, and $|p\rangle$ denotes a
hadron with the momentum $p^\mu=p^+\delta^{\mu+}$. Averages over spin
and color are understood. The above definition is given in the axial gauge
$n\cdot A=0$, where the gauge vector $n$ is assumed to be arbitrary with
$n^2\not= 0$. Though the definition is gauge dependent, physical
observables, such as hadron structure functions and cross sections, are
gauge invariant. It has been shown that the $n$ dependences cancel among
convolution factors of a factorization formula,
{\it i.e.}, among parton distribution functions,
final-state jets, and nonfactorizable soft gluon exchanges, in the
factorization formula for a DIS structure function \cite{L1}. A resummation
can also be performed in the covariant gauge $\partial \cdot A=0$, and the
result is the same as that from the axial gauge \cite{L1}.

The essential step in the CS technique is to derive a
differential equation $p^+d\phi/dp^+=C\phi$ \cite{CS}, where the
coefficient function $C$ contains only single logarithms as shown below,
and can be treated by RG methods. In the axial gauge $n$ appears in the gluon
propagator, $(-i/l^2)N^{\mu\nu}(l)$, with
\begin{equation}
N^{\mu\nu}(l)=g^{\mu\nu}-\frac{n^\mu l^\nu+n^\nu l^\mu}
{n\cdot l}+n^2\frac{l^\mu l^\nu}{(n\cdot l)^2}\;.
\label{gp}
\end{equation}
Because of the scale invariance of $N^{\mu\nu}$ in $n$, $\phi$
depends on $p^+$ through the ratio $(p\cdot n)^2/n^2$, implying that 
the differential operator $d/dp^+$ can be replaced by $d/dn_\alpha$ using a
chain rule,
\begin{equation}
p^+\frac{d}{dp^+}\phi=-\frac{n^2}{v\cdot n}v_{\alpha}
\frac{d}{dn_\alpha}{\phi}\;,
\label{cr}
\end{equation}
where $v=(1,0,{\bf 0})$ is a dimensionless vector along $p$.
The operator $d/dn_\alpha$ applies to $N^{\mu\nu}$, leading to
\begin{eqnarray}
-\frac{n^2}{v\cdot n}v_{\alpha}
\frac{d}{dn_\alpha}N^{\mu\nu}= {\hat v}_{\alpha}
\left(N^{\mu\alpha}l^\nu+N^{\alpha\nu}l^\mu\right)\;,
\label{dgp}
\end{eqnarray}
with the special vertex
\begin{equation}
{\hat v}_{\alpha}=\frac{n^2v_{\alpha}}{v\cdot nn\cdot l}\;.
\end{equation}

The momentum $l^\mu$ ($l^\nu$) is contracted with a vertex the
differentiated gluon attaches, which is then replaced by a special vertex.
For each type of vertices, there exists a Ward identity, which relates a
diagram with the contraction of $l^\mu$ ($l^\nu$) to the difference of two
diagrams \cite{L3}. A pair cancellation then occurs between the contractions
with two adjacent vertices. Summing diagrams with
different differentiated gluons, the special vertex moves to the outer
end of a parton line. We arrive at the derivative,
\begin{equation}
p^+\frac{d}{dp^+}\phi(x,k_T,p^+)=2{\bar \phi}(x,k_T,p^+)\;,
\label{dif}
\end{equation}
described by Fig.~1(a), where the square in the new function
${\bar \phi}$ represents the special vertex ${\hat v}_\alpha$. The
coefficient 2 comes from the equality of the two new functions with
the special vertex on either side of the final-state cut.

To obtain a differential equation of $\phi$, we need to factorize
subdiagrams containing the special vertex out of $\bar\phi$. The
factorization holds in the leading regions of the loop momentum $l$ that
flows through the special vertex. The collinear region of $l$ is not leading
because of the factor $1/(n\cdot l)$ in ${\hat v}_\alpha$ with nonvanishing
$n^2$. Therefore, the leading regions of $l$ are soft and hard, in which the
subdiagrams are factorized from ${\bar \phi}$ into a soft function $K$ and a
hard function $G$, respectively. The remaining part is the original
distribution function $\phi$. That is, $\bar\phi$ is expressed as the
convolution of the functions $K$ and $G$ with $\phi$.

The lowest-order contribution to $K$ from Fig.~1(b) is written as
\begin{equation}
{\bar \phi}_s(x,k_T,p^+)={\bar \phi}_{sv}(x,k_T,p^+)+
{\bar \phi}_{sr}(x,k_T,p^+)\;,
\label{fss}
\end{equation}
with
\begin{eqnarray}
{\bar \phi}_{sv}&=&\left[ig^2C_F\mu^\epsilon
\int\frac{d^{4-\epsilon}l}{(2\pi)^{4-\epsilon}}
N_{\nu\beta}(l)\frac{{\hat v}^\beta v^\nu}{v\cdot l}
\frac{1}{l^2}-\delta K\right]\phi(x,k_T,p^+)\;,
\label{fsv} \\
{\bar \phi}_{sr}&=&ig^2C_F\mu^\epsilon
\int\frac{d^{4-\epsilon}l}{(2\pi)^{4-\epsilon}}
N_{\nu\beta}(l)\frac{{\hat v}^\beta v^\nu}{v\cdot l}
2\pi i\delta(l^2)
\phi(x,|{\bf k}_T+{\bf l}_T|,p^+)\;,
\label{fsr}
\end{eqnarray}
corresponding to virtual and real gluon emissions, respectively.
$C_F=4/3$ is a color factor, and $\delta K$ an additive counterterm.
The ultraviolet pole in Eq.~(\ref{fsv}) is isolated using the dimensional
regularization. To work out the loop integral in Eq.~(\ref{fsr}) explicitly,
we employ the Fourier transformation from $k_T$ space to $b$ space.
The convolution of the subdiagram with $\phi$ in the loop momentum $l_T$ is
then simplified into a product. This is the reason the $k_T$ resummation
should be performed in the space conjugate to $k_T$. The combination of
Eqs.~(\ref{fsv}) and (\ref{fsr}) then gives
\begin{equation}
{\bar\phi}_s(x,b,p^+)=K(1/(b\mu),\alpha_s(\mu))\phi(x,b,p^+)\;,
\end{equation}
with
\begin{eqnarray}
K=ig^2C_F\mu^\epsilon\int\frac{d^{4-\epsilon}l}{(2\pi)^{4-\epsilon}}
N_{\nu\beta}(l)\frac{{\hat v}^\beta v^\nu}{v\cdot l}
\left[\frac{1}{l^2}+2\pi i\delta(l^2)e^{i{\bf l}_T\cdot {\bf b}}\right]
-\delta K\;,
\label{fs1}
\end{eqnarray}
where the factor $\exp(i{\bf l}_T\cdot {\bf b})$ is introduced by the
Fourier transformation.

Note that we have applied the soft approximation to the real gluon emission:
\begin{equation}
\phi(x+l^+/p^+,|{\bf k}_T+{\bf l}_T|,p^+)\approx
\phi(x,|{\bf k}_T+{\bf l}_T|,p^+)\;,
\label{ta}
\end{equation}
when writting Eq.~(\ref{fsr}). This approximation implies that $1/b$
serves as an infrared cutoff for the evaluation of the subdiagrams
shown in Eq.~(\ref{fs1}). Combined with the hard function $G$, which is
characterized by the scale $p^+$, the logarithms $\ln(p^+b)$ are
formed. Hence, Eq.~(\ref{ta}) is associated with the $k_T$ resummation.
We shall show in the next section that the alternative soft approximation
\begin{equation}
\phi(x+l^+/p^+,|{\bf k}_T+{\bf l}_T|,p^+)\approx
\phi(x+l^+/p^+,k_T,p^+)\;,
\label{tta}
\end{equation}
implies the infrared cutoff $p^+/N$ in moment space. Combined with $G$,
the logarithms $\ln(1/N)$ are produced, so that Eq.~(\ref{tta}) is
associated with the threshold resummation.

The lowest-order contribution to $G$ from Fig.~1(c) is given by
\begin{equation}
{\bar\phi}_h(x,b,p^+)=G(xp^+/\mu,\alpha_s(\mu))\phi(x,b,p^+)\;,
\end{equation}
in $b$ space, with
\begin{eqnarray}
G=-ig^2C_F\mu^\epsilon\int\frac{d^{4-\epsilon}l}{(2\pi)^{4-\epsilon}}
N_{\nu\beta}(l)\frac{{\hat v}^\beta}{l^2}
\left[\frac{x\not p-\not l}{(xp- l)^2}\gamma^\nu
+\frac{v^\nu}{v\cdot l}\right]
-\delta G\;,
\label{gpb}
\end{eqnarray}
where $\delta G$ is an additive counterterm. The second term in the above
expression, whose sign is opposite to that of ${\bar\phi}_{sv}$,
is a soft subtraction. This term avoids double counting, and ensures a hard
momentum flow in $G$. At intermediate $x$, $G$ is characterized by a large
scale $p^+$ as stated above.

Using the definition ${\bar\phi}={\bar\phi}_s+{\bar\phi}_h$, Eq.~(\ref{dif})
becomes
\begin{equation}
p^+\frac{d}{dp^+}\phi(x,b,p^+)=2\left[K(1/(b\mu),\alpha_s(\mu))+
G(xp^+/\mu,\alpha_s(\mu))\right]\phi(x,b,p^+)\;,
\label{dph}
\end{equation}
where the sum $K+G$ is the coefficient function $C$ mentioned before. A
straightforward calculation leads Eqs.~(\ref{fs1}) and (\ref{gpb}) to
\begin{eqnarray}
K(1/(b\mu),\alpha_s(\mu))&=&\frac{\alpha_s(\mu)}{\pi}C_F
\ln \frac{1}{b\mu}\;,
\label{kh}\\
G(xp^+/\mu,\alpha_s(\mu))&=&-\frac{\alpha_s(\mu)}{\pi}C_F
\ln\frac{xp^+\nu}{\mu}\;,
\label{gh}
\end{eqnarray}
with $\nu^2=(v\cdot n)^2/|n^2|$ being a gauge factor, where we have 
assumed $n^2<0$ \cite{L1}. Constants of
order unity, which are irrelevant to our discussion, have been neglected.
Equation (\ref{gh}) confirms our argument that $\phi$ depends on $p^+$
via the ratio $(p\cdot n)^2/n^2$. For a detailed derivation of
Eqs.~(\ref{kh}) and (\ref{gh}), refer to \cite{L1}.

The functions $K$ and $G$ possess ultraviolet divergences individually as
indicated by their counterterms. These divergences, both from the virtual
gluon contribution ${\bar \phi}_{sv}$, cancel each other, such that the sum
$K+G$ is RG invariant. The single logarithms $\ln(b\mu)$ and $\ln(p^+/\mu)$,
contained in $K$ and $G$, respectively, are organized by the RG equations
\begin{equation}
\mu\frac{d}{d\mu}K=-\gamma_K=-\mu\frac{d}{d\mu}G\;.
\label{kg}
\end{equation}
The anomalous dimension of $K$, $\lambda_K=\mu d\delta K/d\mu$,
is given, up to two loops, by \cite{BS}
\begin{equation}
\gamma_K=\frac{\alpha_s}{\pi}C_F+\left(\frac{\alpha_s}{\pi}
\right)^2C_F\left[C_A\left(\frac{67}{36}
-\frac{\pi^{2}}{12}\right)-\frac{5}{18}n_{f}\right]\;,
\label{lk}
\end{equation}
with $n_{f}$ being the number of quark flavors, and $C_A=3$ a color factor.

Solving Eq.~(\ref{kg}), we have
\begin{eqnarray}
& &K(1/(b\mu),\alpha_s(\mu))+G(xp^+/\mu,\alpha_s(\mu))
\nonumber\\
& &=K(1,\alpha_s(1/b))+G(1,\alpha_s(xp^+))
-\int_{1/b}^{xp^+}\frac{d\mu}{\mu}\gamma_K(\alpha_s(\mu))\;,
\nonumber\\
& &=-\int_{1/b}^{xp^+}\frac{d\mu}{\mu}\gamma_K(\alpha_s(\mu))\;,
\label{skg}
\end{eqnarray}
where the initial condition $K(1,\alpha_s(1/b))$ vanishes, and $\ln\nu$
in $G(1,\alpha_s(xp^+))$ has been dropped, since the gauge factor will be
cancelled as computing physical quantities. Inserting the above expression
into Eq.~(\ref{dph}), we obtain the solution
\begin{eqnarray}
\phi(x,b,p^+)=\Delta_k(b,xp^+)\phi(x)
\label{sph}
\end{eqnarray}
with the Sudakov exponential from the $k_T$ resummation,
\begin{equation}
\Delta_k(b,xp^+)=\exp\left[-\int_{1/b}^{xp^+}\frac{d p}{p}
\int_{1/b^2}^{p^2}\frac{d\mu^2}{\mu^2}
\gamma_{K}(\alpha_s(\mu))\right]\;.
\label{fb}
\end{equation}
We have set the upper bound of the variable $p$ to $xp^+$, which
corresponds to the final condition $\phi(x,b,p^+)$, and the lower bound
to $1/b$, such that the initial condition $\phi(x)$ does not contain
the logarithms $\ln(p^+b)$. This statement will become essential,
when Eq.~(\ref{fb}) is compared with the exponential from the threshold
resummation for $\phi$.

\vskip 1.0cm

\centerline{\large \bf III. THRESHOLD RESUMMATION}
\vskip 0.5cm

In this section we derive, using the CS formalsim, the threshold
resummation for the quark distribution funciton,
\begin{equation}
\phi(x,p^+)=\int\frac{dy^-}{2\pi}e^{-ix p^+y^-}
\langle p| {\bar q}(y^-)\frac{1}{2}\gamma^+q(0)|p\rangle\;,
\label{det}
\end{equation}
which is obtained by integrating Eq.~(\ref{dep}) over $k_T$. Comparing each
step of the derivation, the relation of the threshold
resummation to the $k_T$ resummation will be clear.
The argument $p^+$ represents the logarithms $\ln(1-x)p^+$.
According to the same reasoning, $\phi$ depends on $p^+$ through the ratio
$(p\cdot n)^2/n^2$, and the chain rule in Eq.~(\ref{cr})
holds. Following Eqs.~(\ref{dgp}) and (\ref{dif}), we have the
factorization of ${\bar \phi}$ into the convolution of subdiagrams
containing the special vertex with $\phi$. Similarly, subdiagrams with
soft and hard loop momentum flows
are absorbed into the functions $K$ and $G$, respectively.

The lowest-order contribution to $K$ from Fig.~1(b) is written as
\begin{equation}
{\bar \phi}_s(x,p^+)={\bar \phi}_{sv}(x,p^+)+
{\bar \phi}_{sr}(x,p^+)\;,
\label{tss}
\end{equation}
with
\begin{eqnarray}
{\bar \phi}_{sv}&=&\left[ig^2C_F\mu^\epsilon
\int\frac{d^{4-\epsilon}l}{(2\pi)^{4-\epsilon}}
N_{\nu\beta}(l)\frac{{\hat v}^\beta v^\nu}{v\cdot l}
\frac{1}{l^2}-\delta K\right]\phi(x,p^+)\;,
\label{tsv} \\
{\bar \phi}_{sr}&=&ig^2C_F\mu^\epsilon
\int\frac{d^{4-\epsilon}l}{(2\pi)^{4-\epsilon}}
N_{\nu\beta}(l)\frac{{\hat v}^\beta v^\nu}{v\cdot l}
2\pi i\delta(l^2)\phi(x+l^+/p^+,p^+)\;,
\label{tsr}
\end{eqnarray}
corresponding to virtual and real gluon emissions, respectively. The
virtual gluon contribution is the same as that in Eq.~(\ref{fsv}),
and Eq.~(\ref{tsr}) is the consequence of the soft approximation
in Eq.~(\ref{tta}). If retaining transverse degrees of freedom of a
parton at the beginning, Eq.~(\ref{tta}) allows us to integrate out the
$k_T$ dependences of ${\bar\phi}_{sr}$ and of $\phi$, and we still arrive
at Eq.~(\ref{tsr}).

Inserting the identities
\begin{eqnarray}
\int_x^1d\xi \delta(\xi-x)=1\;,\;\;\;\;
\int_x^1d\xi \delta(\xi-x-l^+/p^+)=1\;,
\label{delt}
\end{eqnarray}
into Eqs.~(\ref{tsv}) and (\ref{tsr}), respectively, Eq.~(\ref{tss})
is reexpressed as
\begin{eqnarray}
{\bar \phi}_{s}(x,p^+)=\int_x^1\frac{d\xi}{\xi}
K\left(\left(1-\frac{x}{\xi}\right)\frac{p^+}{\mu},\alpha_s(\mu)\right)
\phi(\xi,p^+)\;,
\label{tss1}
\end{eqnarray}
with
\begin{eqnarray}
K&=&ig^2C_F\mu^\epsilon\int\frac{d^{4-\epsilon}l}{(2\pi)^{4-\epsilon}}
N_{\nu\beta}(l)\frac{{\hat v}^\beta v^\nu}{v\cdot l}
\left[\frac{\delta(1-x/\xi)}{l^2}\right.
\nonumber\\
& &\left.+2\pi i\delta(l^2)\delta\left(1-\frac{x}{\xi}-\frac{l^+}{p^+}
\right)\right]-\delta K\delta\left(1-\frac{x}{\xi}\right)\;.
\label{kss}
\end{eqnarray}
To obtain the above expression, we have adopted the approximation
\begin{eqnarray}
\delta\left(1-\frac{x}{\xi}-\frac{l^+}{\xi p^+}\right)
\approx\delta\left(1-\frac{x}{\xi}-\frac{l^+}{p^+}\right)\;,
\end{eqnarray}
which is appropriate in the considered region with $x\to 1$.

To work out the $\xi$ integration explicitly, we employ a Mellin
transformation from momentum fraction ($x$) space to moment ($N$)
space,
\begin{eqnarray}
{\bar\phi}_{s}(N,p^+)&\equiv &\int_0^1 dxx^{N-1}{\bar\phi}_{s}(x,p^+)\;,
\nonumber\\
&=&K(p^+/(N\mu),\alpha_s(\mu))\phi(N,p^+)\;.
\end{eqnarray}
with
\begin{eqnarray}
K(p^+/(N\mu),\alpha_s(\mu))=\int_0^1 dzz^{N-1}
K((1-z)p^+/\mu,\alpha_s(\mu))\;.
\label{kmt}
\end{eqnarray}
The convolution between $K$ and $\phi$ is then simplified into
a product. This is the reason the threshold resummation should be performed
in moment space. The Mellin transformation for the threshold resummation
and the Fourier transformation for the $k_T$ resummation then play the same
role. As shown in the Appendix, $K$ is given by
\begin{eqnarray}
K(p^+/(N\mu),\alpha_s(\mu))
&=&\frac{\alpha_s(\mu)}{\pi}C_F\left(\int_0^1 dz\frac{z^{N-1}-1}{1-z}
+\ln\frac{p^+\nu}{\mu}\right)\;,
\nonumber\\
&=&\frac{\alpha_s(\mu)}{\pi}C_F\ln\frac{p^+\nu}{N\mu}\;,
\label{ts1}
\end{eqnarray}
with the same counterterm $\delta K$. To derive the second expression, we
have identified the integral over $z$ as $\ln(1/N)$, which is valid up to
corrections suppressed by $1/N$. Hence, soft real gluon emissions
produce the logarithms $\ln(p^+/N)$ as stated before.

The lowest-order contribution to the hard function $G$ from Fig.~1(c) is
written as
\begin{equation}
{\bar\phi}_h(N,p^+)=G(p^+/\mu,\alpha_s(\mu))\phi(N,p^+)\;,
\label{gtr}
\end{equation}
in $N$ space, where the expressions of $G$ have been given in
Eqs.~(\ref{gpb}) and (\ref{gh}) with $x=1$. In conclusion, the functional
forms of $K$ and $G$ in the threshold resummation are the same as
Eqs.~(\ref{kh}) and (\ref{gh}), but with the scales $1/b$ replaced
by $p^+\nu/N$ and $xp^+$ by $p^+$, respectively.

Using ${\bar\phi}={\bar\phi}_s+{\bar\phi}_h$, we have the differential
equation
\begin{equation}
p^+\frac{d}{dp^+}\phi(N,p^+)=2\left[K(p^+/(N\mu),\alpha_s(\mu))+
G(p^+/\mu,\alpha_s(\mu))\right]\phi(N,p^+)\;.
\label{tph}
\end{equation}
Again, $K$ and $G$ contain ultraviolet divergences
individually, but their sum $K+G$ is RG invariant.
The RG solution of $K+G$ is given by
\begin{eqnarray}
K(p^+/(N\mu),\alpha_s(\mu))+G(p^+/\mu,\alpha_s(\mu))=
-\int_{p^+/N}^{p^+}\frac{d\mu}{\mu}\gamma_K(\alpha_s(\mu))\;.
\label{tkg}
\end{eqnarray}
To sum $\ln(1/N)$ by means of Eq.~(\ref{tph}), we make the replacement
\begin{equation}
p^+\frac{d\phi}{dp^+}=\frac{p^+}{N}\frac{d\phi}{d (p^+/N)}\;.
\label{ren}
\end{equation}
That is, the characteristic scale $p^+$ of $G$ is frozen, when solving
Eq.~(\ref{tph}). We then obtain
\begin{eqnarray}
\phi(N,p^+)=\Delta_t(N,p^+)\phi^{(0)}\;,
\label{pht}
\end{eqnarray}
with the exponential from the threshold resummation,
\begin{eqnarray}
\Delta_t(N,p^+)=\exp\left[-\int_{p^+}^{p^+/N}\frac{d p}{p}
\int_{p^2}^{p^{+2}}\frac{d\mu^2}{\mu^2}
\gamma_{K}(\alpha_s(\mu))\right]\;,
\label{fbt}
\end{eqnarray}
which exhibits enhancement.

We have set the upper bound of the variable $p$ to $p^+/N$, which
corresponds to the final condition $\phi(N,p^+)$ with $p^+/p=N$, and the
lower bound to $p^+$, such that the initial condition $\phi^{(0)}$ does not
contain $\ln(1/N)$ because of $p^+/p=1$. In this way the
$N$ dependence of $\phi$ is grouped into the exponential $\Delta_t$.
Contrary to Eq.~(\ref{fb}), the lower bound of $p$ is larger than the upper
bound in $\Delta_t$. Hence, the $k_T$ resummation gives the evolution of
a parton distribution function from the small scale $1/b$ to the large scale
$p^+$, while the threshold resummation gives the evolution from the large
$p^+$ to the small $p^+/N$. Since the directions of the double-logarithm
evolutions are opposite, their resummation effects are also opposite.

Employing the variable change $p=(1-z)p^+$ and $\mu=\sqrt{\lambda} p^+$,
Eq.~(\ref{fbt}) becomes
\begin{eqnarray}
\Delta_t(N,p^+)=\exp\left[\int_{0}^{1-1/N}\frac{d z}{1-z}
\int_{(1-z)^2}^{1}\frac{d\lambda}{\lambda}
\gamma_{K}(\alpha_s(\sqrt{\lambda} p^+))\right]\;.
\label{fbt1}
\end{eqnarray}
It can be easily justified that the above expression is equivalent to
\begin{eqnarray}
\Delta_t(N,p^+)=\exp\left[-\int_{0}^{1}dz\frac{z^{N-1}-1}{1-z}
\int_{(1-z)^2}^{1}\frac{d\lambda}{\lambda}
\gamma_{K}(\alpha_s(\sqrt{\lambda} p^+))\right]\;,
\label{fbt2}
\end{eqnarray}
up to $O(1/N)$ corrections, which has appeared in \cite{S}.

\vskip 1.0cm

\centerline{\large \bf IV. THRESHOLD RESUMMATION FOR A JET}
\vskip 0.5cm

As stated in the Introduction, the scattered quark in DIS carries the
momentum $p'=\xi p+q=((\xi-x)p^+,q^-,{\bf 0})$, where $\xi p$ is the initial
quark momentum and $q$ the photon momentum. In the threshold region with
$x\to \xi\to 1$, $p'$ possesses a large minus component $p'^-=q^-$ but a
small invariant $p'^2=2(\xi-x)p\cdot q\equiv (\xi-x)s$. This scattered quark
produces a jet of particles in this region, to which involved radiative
corrections contain both collinear divergences from $l$ parallel to $p'$ and
soft divergences from $l\to 0$. Because of the cancellation of
soft divergences, these corrections are mainly collinear, and absorbed
into a jet function $J$.
This jet function $J$ depends on the ratio
$p'^2/s=\xi-x$, and more precisely, on the momentum fraction $w=1-x/\xi$ in
a factorization formula. For convenience, we parametrize the scattered
quark momentum as
$p'=(wp^{\prime-},p^{\prime-},{\bf 0})$, which can be achieved by
choosing a frame with $p^+=q^-=p^{\prime-}$.

We perform the threshold resummation for $J(w,p^{\prime-})$ in the CS
framework. It will be confirmed below that the large scale
$p^{\prime-}$ appears in the ratio
\begin{equation}
\frac{(n\cdot v')^2}{n^2p^{\prime-2}}\;,
\label{rj}
\end{equation}
in the case with nonvanishing jet invariant mass, where $v'=(w,1,{\bf 0})$
is a dimensionless vector along $p'$. The above ratio differs from
$(p\cdot n)^2/n^2$ in the massless case for the quark distribution function,
which has been discussed in the previous section. We shall demonstrate that
this difference turns the threshold resummation from enhancement for a
parton distribution function to suppression for a jet function.
Note that Eq.~(\ref{rj}) does not approach $(p'\cdot n)^2/n^2$ as
$w\to 0$. This explains why the threshold resummations for a jet function
and for a parton distribution function do not coincide with each other in
the large $N$ (small $w$ and large $x$) limit.

The chain rule for $J$, corresponding to Eq.~(\ref{cr}) for $\phi$,
becomes
\begin{equation}
-p^{\prime-}\frac{d}{dp^{\prime-}}J=
-\frac{n^2}{v'\cdot n}v'_\alpha\frac{d}{dn_\alpha}J\;.
\label{cj}
\end{equation}
Obviously, the extra minus sign on the left-hand side of the above
equation will lead to suppression.
Following the same procedures as in Sec. III, we obtain the derivative
\begin{equation}
-p^{\prime-}\frac{d}{dp^{\prime-}}J(w,p^{\prime-})
=2{\bar J}(w,p^{\prime-})\;,
\label{dij}
\end{equation}
where the new function ${\bar J}$ contains the special vertex
\begin{equation}
{\hat v}'_\alpha=\frac{n^2v'_\alpha}{v'\cdot nn\cdot l}\;.
\label{nvp}
\end{equation}
The coefficient 2, again, comes from the equality of the two new functions
with the special vertex on either side of the final-state cut.

Similarly, the leading regions of the loop momentum $l$ flowing through
the special vertex are soft and hard, in which subdiagrams containing
the special vertex are factorized out of ${\bar J}$ into a soft function
$K$ and a hard function $G$, respectively. The lowest-order subdiagrams for
$K$ and for $G$ are basically the same as those in Figs.~1(b) and 1(c). The
contribution from Fig.~1(b) is written as
\begin{equation}
{\bar J}_s(w,p^{\prime-})={\bar J}_{sv}(w,p^{\prime-})
+{\bar J}_{sr}(w,p^{\prime-})\;,
\label{jss}
\end{equation}
with
\begin{eqnarray}
{\bar J}_{sv}&=&\left[ig^2C_F\mu^\epsilon
\int\frac{d^{4-\epsilon}l}{(2\pi)^{4-\epsilon}}
N_{\nu\beta}(l)\frac{{\hat v}^{\prime\beta} v^{\prime\nu}}{v'\cdot l}
\frac{1}{l^2}-\delta K\right]J(w,p^{\prime-})\;,
\label{jsv} \\
{\bar J}_{sr}&=&ig^2C_F\int\frac{d^{4-\epsilon}l}{(2\pi)^{4-\epsilon}}
N_{\nu\beta}(l)\frac{{\hat v}^{\prime\beta} v^{\prime\nu}}{v'\cdot l}
2\pi i\delta(l^2)J((p'-l)^2/s,p^{\prime-})\;,
\label{jsr}
\end{eqnarray}
corresponding to virtual and real gluon emissions, respectively. The
virtual gluon contribution is the same as in Eq.~(\ref{tsv}) except for
the substitution of the vector $v'$ for $v$, and thus the counterterm
is also $\delta K$. The first argument of $J$ associated with the real gluon
emission,
\begin{equation}
\frac{(p'-l)^2}{s}=w-\frac{l^+}{p^{\prime-}}-\frac{wl^-}{p^{\prime-}}\;,
\end{equation}
differs from the corresponding one in Eq.~(\ref{tsr}).

Inserting the identities
\begin{eqnarray}
\int_0^wdy \delta(w-y)=1\;,\;\;\;\;
\int_0^wdy \delta\left(w-y-\frac{l^+}{p^{\prime-}}-\frac{wl^-}{p^{\prime-}}
\right)=1\;,
\end{eqnarray}
into Eqs.~(\ref{jsv}) and (\ref{jsr}), respectively, Eq.(\ref{jss})
is reexpressed as
\begin{eqnarray}
{\bar J}_s(w,p^{\prime-})=
\int_0^w\frac{dy}{1-y}K\left(\frac{w-y}{1-y}\frac{p^{\prime-}}{\mu},
\alpha_s(\mu)\right)J(y,p^{\prime-})\;,
\label{jss1}
\end{eqnarray}
with
\begin{eqnarray}
K&=&ig^2C_F\mu^\epsilon\int\frac{d^{4-\epsilon}l}{(2\pi)^{4-\epsilon}}
N_{\nu\beta}(l)\frac{{\hat v}^\beta v^\nu}{v\cdot l}
\left[\frac{1}{l^2}\delta\left(\frac{w-y}{1-y}\right)\right.
\nonumber\\
& &\left.+2\pi i\delta(l^2)\delta\left(\frac{w-y}{1-y}
-\frac{l^+}{p^{\prime-}}-\frac{wl^-}{p^{\prime-}}\right)\right]
-\delta K\delta\left(\frac{w-y}{1-y}\right)\;.
\label{kjss}
\end{eqnarray}
To obtain the above expression, we have adopted the approximation
\begin{eqnarray}
\delta\left(\frac{w-y}{1-y}
-\frac{l^+}{(1-y)p^{\prime-}}-\frac{wl^-}{(1-y)p^{\prime-}}\right)
\approx\delta\left(\frac{w-y}{1-y}
-\frac{l^+}{p^{\prime-}}-\frac{wl^-}{p^{\prime-}}\right)\;,
\end{eqnarray}
which is appropriate in the considered region with $w\to 0$.

To work out the $y$ integration explicitly, we apply a Mellin transformation
from momentum fraction ($w$) space to moment ($N$) space,
\begin{eqnarray}
{\bar J}_s(N,p^{\prime-})&=&\int_0^1 dw (1-w)^{N-1}
{\bar J}_s(w,p^{\prime-})\;,
\nonumber\\
&=&\int_0^1 \frac{dy}{1-y}\int_y^1 dw(1-w)^{N-1}
K\left(\frac{w-y}{1-y}\frac{p^{\prime-}}{\mu},
\alpha_s(\mu)\right)J(y,p^{\prime-})\;,
\nonumber\\
& &
\label{jme}
\end{eqnarray}
Using the variable change $w=z(1-y)+y$, the above expression
reduces to
\begin{equation}
{\bar J}_s(N,p^{\prime-})=K(p^{\prime-}/(N\mu),\alpha_s(\mu))
J(N,p^{\prime-})\;,
\label{kjn}
\end{equation}
with
\begin{eqnarray}
K(p^{\prime-}/(N\mu),\alpha_s(\mu))=\int_0^1 dz (1-z)^{N-1}
K(zp^{\prime-}/\mu,\alpha_s(\mu))\;.
\end{eqnarray}
Replacing the variable $z$ by $1-z$, and following the procedures in the
Appendix, the result of $K$ is 
\begin{eqnarray}
K(p^{\prime-}/(N\mu),\alpha_s(\mu))
&=&\frac{\alpha_s(\mu)}{\pi}C_F\left(\int_0^1 dz\frac{z^{N-1}-1}{1-z}
+\ln\frac{p^{\prime-}}{\nu'\mu}\right)\;,
\nonumber\\
&=&\frac{\alpha_s(\mu)}{\pi}C_F\ln\frac{p^{\prime-}}{N\nu'\mu}\;,
\label{js1}
\end{eqnarray}
with the gauge factor $\nu'^2=(n\cdot v')^2/|n^2|$, which confirms that
$J$ depends on $p'^-$ via the ratio $(n\cdot v')^2/(n^2p^{\prime-2})$ in
the massive case.

An alternative Mellin transformation adopted in \cite{S} is
\begin{eqnarray}
{\bar J}_s(N,p^{\prime-})&=&\int_0^\infty dw e^{-Nw}
{\bar J}_s(w,p^{\prime-})\;,
\nonumber\\
&=&\int_0^\infty dy\int_y^\infty dw e^{-Nw}
K((w-y)p^{\prime-}/\mu,\alpha_s(\mu))J(y,p^{\prime-}).
\label{amj0}
\end{eqnarray}
Employing the variable change $w=z+y$, the above expression reduces to
Eq.~(\ref{kjn}), but with
\begin{eqnarray}
K(p^{\prime-}/(N\mu),\alpha_s(\mu))=\int_0^\infty dz e^{-Nz}
K(zp^{\prime-}/\mu,\alpha_s(\mu))\;.
\label{amj}
\end{eqnarray}
It is easy to justify the equivalence between the transformation in
Eq.~(\ref{amj0}) and that in Eq.~(\ref{jme}).

The lowest-order contribution to the hard function $G$ from Fig.~1(c)
is given by
\begin{equation}
{\bar J}_h(w,p^{\prime-})=G(\sqrt{w}p^{\prime-}/\mu,\alpha_s(\mu))
J(w,p^{\prime-})\;,
\label{ht}
\end{equation}
with the hard function
\begin{eqnarray}
G=-ig^2C_F\mu^\epsilon\int\frac{d^{4-\epsilon}l}{(2\pi)^{4-\epsilon}}
N_{\nu\beta}(l)\frac{{\hat v}^{\prime\beta}}{l^2}
\left[\frac{\not p'-\not l}{(p'- l)^2}\gamma^\nu
-\frac{2p^{\prime\nu}}{p'^2-2p'\cdot l}\right]
-\delta G\;.
\label{gtj}
\end{eqnarray}
Note the modified eikonal approximation for the quark propagator
$(p'-l)^2\approx p'^2-2p'\cdot l$, since we are considering the massive
case with $p'^2\not=0$. As shown in the Appendix, the expression
for $G$ is 
\begin{eqnarray}
G=-\frac{\alpha_s(\mu)}{\pi}C_F
\ln\frac{\sqrt{w}p^{\prime-}}{\nu'\mu}\;.
\label{gj}
\end{eqnarray}
As applying the Mellin transformation in Eq.~(\ref{amj0}) to both sides of
Eq.~(\ref{ht}), we simply replace the momentum fraction $w$ in $G$ by $1/N$
because of
\begin{eqnarray}
\int_0^\infty dw e^{-Nw}\ln w{\bar J}(w,p^{\prime-})
= \ln\frac{1}{N}\int_0^\infty dw e^{-Nw}{\bar J}(w,p^{\prime-})\;,
\end{eqnarray}
which is valid up to $O(1/N)$ corrections. Equation (\ref{ht}) then
leads to 
\begin{equation}
{\bar J}_h(N,p^{\prime-})=G(p^{\prime-}/(\sqrt{N}\mu),\alpha_s(\mu))
J(N,p^{\prime-})\;.
\label{htn}
\end{equation}

Using ${\bar J}={\bar J}_s+{\bar J}_h$, Eq.~(\ref{dij}) becomes
\begin{eqnarray}
-p^{\prime-}\frac{d}{dp^{\prime-}}J(N,p^{\prime-})
&=&2[K(p^{\prime-}/(N\mu),\alpha_s(\mu))+G(p^{\prime-}/(\sqrt{N}\mu),
\alpha_s(\mu))]J(N,p^{\prime-})\;.
\label{jij}
\end{eqnarray}
The RG analysis gives
\begin{eqnarray}
K(p^{\prime-}/(N\mu),\alpha_s(\mu))+G(p^{\prime-}/(\sqrt{N}\mu),
\alpha_s(\mu))
=-\frac{1}{2}\int_{p^{\prime-2}/N^2}^{p^{\prime-2}/N}
\frac{d\mu^2}{\mu^2}\gamma_K(\alpha_s(\mu))\;.
\label{jkg}
\end{eqnarray}
Solving the differential equation (\ref{jij}) under a modification of the
derivative similar to Eq.~(\ref{ren}), we derive
\begin{eqnarray}
J(N,p^{\prime-})=\Delta'_t(N,p^{\prime-})J^{(0)}\;,
\label{phj}
\end{eqnarray}
with the exponential from the threshold resummation,
\begin{eqnarray}
\Delta'_t(N,p^{\prime-})=\exp\left[\int_{p^{\prime-}}^{p^{\prime-}/N}
\frac{d p}{p}\int_{p^2/N^2}^{p^{\prime-}p/N}\frac{d\mu^2}{\mu^2}
\gamma_{K}(\alpha_s(\mu))\right]\;.
\label{fbj}
\end{eqnarray}
We have set the upper bound of the variable $p$ to $p^{\prime-}/N$ and the
lower bound to $p^{\prime-}$, which correspond to the final condition
$J(N,p^{\prime-})$ and to the initial condition $J^{(0)}$, respectively.

At last, adopting the variable changes $p=(1-z)p^{\prime-}$ and
$\mu=\sqrt{\lambda} p^{\prime-}$, Eq.~(\ref{fbj}) is reexpressed as
\begin{eqnarray}
\Delta'_t(N,p^{\prime-})=\exp\left[-\int_{0}^{1-1/N}\frac{d z}{1-z}
\int_{(1-z)^2}^{1-z}\frac{d\lambda}{\lambda}
\gamma_{K}(\alpha_s(\sqrt{\lambda} p^{\prime-}))\right]\;,
\label{fbj1}
\end{eqnarray}
which is equivalent to \cite{S}
\begin{eqnarray}
\Delta'_t(N,p^{\prime})=\exp\left[\int_{0}^{1}dz\frac{z^{N-1}-1}{1-z}
\int_{(1-z)^2}^{1-z}\frac{d\lambda}{\lambda}
\gamma_{K}(\alpha_s(\sqrt{\lambda} p^{\prime-}))\right]\;,
\label{fbj2}
\end{eqnarray}
up to corrections suppressed by $1/N$.

\vskip 1.0cm

\centerline{\large \bf V. UNIFIED RESUMMATION}
\vskip 0.5cm

In the previous sections we have shown that both the $k_T$ and threshold
resummations can be reproduced in the CS framework by introducing the
transverse and longitudinal cutoffs for soft real gluon emissions,
respectively. It is then possible to derive a unification of the two
resummations for a parton distribution function in the same formalism by
keeping both cutoffs. Consider the quark distribution function
defined by Eq.~(\ref{dep}). Adopting the similar reasoning, we have
Eq.~(\ref{dif}), and the factorization of ${\bar \phi}$ into the convolution
of subdiagrams containing the special vertex with $\phi$: the subdiagrams
with soft and hard loop momentum flows are absorbed into the functions $K$
and $G$, respectively.

The lowest-order contribution to ${\bar \phi}_s$ from Fig.~1(b)
is written as Eq.~(\ref{fss}), with ${\bar \phi}_{sv}(x,k_T,p^+)$ the
same as Eq.~(\ref{fsv}) and
\begin{eqnarray}
{\bar \phi}_{sr}=ig^2C_F\mu^\epsilon
\int\frac{d^{4-\epsilon}l}{(2\pi)^{4-\epsilon}}
N_{\nu\beta}(l)\frac{{\hat v}^\beta v^\nu}{v\cdot l}
2\pi i\delta(l^2)
\phi(x+l^+/p^+,|{\bf k}_T+{\bf l}_T|,p^+)\;.
\label{fsru}
\end{eqnarray}
Note that both the $l^+$ and $l_T$ dependences of $\phi$ in the integrand
of ${\bar \phi}_{sr}$ have been retained. To work out the $l_T$ integration
explicitly, we employ the Fourier transformation from $k_T$ space
to $b$ space as shown in Eq.~(\ref{fs1}). Following the
derivation of Eq.~(\ref{kss}), ${\bar\phi}_s$ becomes
\begin{eqnarray}
{\bar \phi}_s(x,b,p^+)=\int_x^1\frac{d\xi}{\xi}
K\left(\left(1-\frac{x}{\xi}\right)\frac{p^+}{\mu},1/(b\mu),
\alpha_s(\mu)\right)\phi(\xi,b,p^+)\;,
\label{tss1u}
\end{eqnarray}
with
\begin{eqnarray}
K&=&ig^2C_F\mu^\epsilon\int\frac{d^{4-\epsilon}l}{(2\pi)^{4-\epsilon}}
N_{\nu\beta}(l)\frac{{\hat v}^\beta v^\nu}{v\cdot l}
\left[\frac{\delta(1-x/\xi)}{l^2}\right.
\nonumber\\
& &\left.+2\pi i\delta(l^2)\delta\left(1-\frac{x}{\xi}-\frac{l^+}{p^+}
\right)e^{i{\bf l}_T\cdot {\bf b}}\right]
-\delta K\delta\left(1-\frac{x}{\xi}\right)\;.
\label{kssu}
\end{eqnarray}
Further applying the Mellin transformation to Eq.~(\ref{tss1u}), we obtain
\begin{eqnarray}
{\bar\phi}_{s}(N,b,p^+)
=K(p^+/(N\mu),1/(b\mu),\alpha_s(\mu))\phi(N,b,p^+)\;,
\end{eqnarray}
with 
\begin{eqnarray}
K(p^+/(N\mu),1/(b\mu),\alpha_s(\mu))=\int_0^1 dzz^{N-1}
K((1-z)p^+/\mu,1/(b\mu),\alpha_s(\mu)).
\label{kmtu}
\end{eqnarray}
The convolutions between $K$ and $\phi$ in $l^+$ and $l_T$ are then
completely simplified into a multiplication under the Mellin and Fourier
transformations, respectively.

It is easy to work out the integration in Eq.~(\ref{kmtu}), and the result
is
\begin{eqnarray}
K(p^+/(N\mu),1/(b\mu),\alpha_s(\mu))
=\frac{\alpha_s(\mu)}{\pi}C_F\left[\ln\frac{1}{b\mu}
-K_0\left(\frac{2\nu p^+b}{N}\right)\right]\;,
\label{uk}
\end{eqnarray}
$K_0$ being the modified Bessel function. We examine
the large $p^+b$ and $N$ limits of the above expression. For $p^+b\gg N$, 
we have $K_0\to 0$ and 
\begin{equation}
K\to\frac{\alpha_s}{\pi}C_F\ln\frac{1}{b\mu}\;,
\label{lpb}
\end{equation}
which is the soft function for the $k_T$ resummation in
Eq.~(\ref{kh}). For $N\gg p^+b$, we have $K_0\approx -\ln(\nu p^+b/N)$ and 
\begin{equation}
K\to\frac{\alpha_s}{\pi}C_F\ln\frac{\nu p^+}{N\mu}\;,
\label{ln}
\end{equation}
which is the soft function for the threshold resummation in Eq.~(\ref{ts1}).
Hence, Eq.~(\ref{uk}) is indeed appropriate for the unification of the $k_T$
and threshold resummations.

The lowest-order contribution to $G$ from Fig.~1(c) is the same as in
Eq.~(\ref{gtr}). Using ${\bar\phi}={\bar\phi}_s+{\bar\phi}_h$,
Eq.~(\ref{dif}) becomes
\begin{eqnarray}
p^+\frac{d}{dp^+}\phi(N,b,p^+)=2\left[K(p^+/(N\mu),1/(b\mu),
\alpha_s(\mu))+G(p^+/\mu,\alpha_s(\mu))\right]\phi(N,b,p^+)\;.
\label{dphu}
\end{eqnarray}
As solving the RG equations (\ref{kg}), we allow the variable $\mu$ evolves
from the characteristic scale of $K$ to the scale of $G$. Equation (\ref{uk})
implies the characteristic scale of order
\begin{equation}
\frac{1}{b}\exp\left[-K_0\left(\frac{p^+b}{N}\right)\right]\;,
\label{usc}
\end{equation}
for the unified resummation. We discuss the
cases for $p^+b\gg N$ and for $N\gg p^+b$ first, which will help the
derivation of the unified resummation. The solution of $K+G$ is written as 
\begin{eqnarray}
& &K(p^+/(N\mu),1/(b\mu),\alpha_s(\mu))+G(p^+/\mu,\alpha_s(\mu))
\nonumber\\
&=&-\int_{1/b}^{p^+}\frac{d\mu}{\mu}\gamma_K(\alpha_s(\mu))\;,
\;\;\;\;{\rm for}\;\;p^+b\gg N\;,
\nonumber\\
& &-\int_{p^+/N}^{p^+}\frac{d\mu}{\mu}\gamma_K(\alpha_s(\mu))\;,
\;\;\;\;{\rm for}\;\;N\gg p^+b\;,
\label{skgu}
\end{eqnarray}
indicating that the distribution function
$\phi(N,b,p^+)$ involves $\ln(p^+b)$ and $\ln(1/N)$ in
the $p^+b\to \infty$ and $N\to \infty$ limits, respectively.
This can be easily understood by ignoring the variation of
$\gamma_K$, and performing the $\mu$ integration directly.

Inserting Eq.~(\ref{skgu}) into (\ref{dphu}), we obtain the solutions 
\begin{eqnarray}
\phi(N,b,p^+)&=&\exp\left[-2\int_{\exp[-K_0(p^+b/N)]/b}^{p^+}\frac{d p}{p}
\int_{1/b}^{p}\frac{d\mu}{\mu}
\gamma_{K}(\alpha_s(\mu))\right]\phi^{(0)}\;,
\label{fb2}\\
\phi(N,b,p^+)&=&\exp\left[-2\int_{\exp[-K_0(p^+b/N)]/b}^{p^+}\frac{dp}{p}
\int_{p^+}^{p}\frac{d\mu}{\mu}
\gamma_{K}(\alpha_s(\mu))\right]\phi^{(0)}\;,
\label{ft4}
\end{eqnarray}
for $p^+b\gg N$ and $N\gg p^+b$, respectively. We have
chosen the characteristic scale in Eq.~(\ref{usc}) as the lower bound of
the variable $p$. To unify the above expressions, we replace the lower
bounds of $\mu$ by
\begin{equation}
\frac{1}{b}\exp[-K_0(p^+b)]\;,
\end{equation}
that is motivated by Eq.~(\ref{usc}). At last, the unified resummation is
given by
\begin{equation}
\phi(N,b,p^+)=\Delta_u(N,b,p^+)\phi^{(0)},
\label{fb3}
\end{equation}
with the exponential
\begin{equation}
\Delta_u(N,b,p^+)=
\exp\left[-2\int_{\exp[-K_0(p^+b/N)]/b}^{p^+}\frac{d p}{p}
\int_{\exp[-K_0(p^+b)]/b}^{p}\frac{d\mu}{\mu}
\gamma_{K}(\alpha_s(\mu))\right],
\label{fb4}
\end{equation}
which is appropriate for arbitrary $p^+b$ and $N$. It is easy to justify
that Eq.~(\ref{fb4}) approaches the $k_T$ resummation in Eq.~(\ref{fb})
as $b\to\infty$, and approaches the threshold resummation in
Eq.~(\ref{fbt}) as $b\to 0$.

\vskip 1.0cm
\centerline{\large \bf VI. CONCLUSION}
\vskip 0.5cm

In this paper we have explored the relation between the $k_T$ and threshold
resummations, and discussed the difference between their applications to a
parton distribution function and to a jet function in the CS framework. It
has been understood that the summations of the logarithms $\ln^2(p^+b)$ and
$\ln^2(1/N)$ are determined by the soft approximations for real gluon
emission in Eqs.~(\ref{ta}) and (\ref{tta}), respectively. Suppression
from the $k_T$ resummation, as indicated by $\Delta_k$ in Eq.~(\ref{fb}),
and  enhancement from the threshold resummation, as indicated by
$\Delta_t$ in Eq.~(\ref{fbt2}), for a parton distribution function is
the consequence of the opposite directions of their double-logarithm
evolutions. While the switch of the threshold resummation from 
enhancement for a parton distribution function to suppression for a jet
function is attributed to the change from a massless case to a massive case.
The $k_T$ resummation for a jet function remains suppressive as shown in
Eq.~(\ref{fj0}), since the jet momentum and the parton momentum are in the
same form. Hence, we do not repeat the derivation here.

We have also unified the $k_T$ and threshold resummations 
by keeping simultaneously the longitudianl and
transverse loop momenta in the parton distribution function
$\phi(x+l^+/p^+,|{\bf k}_T+{\bf l}_T|,p^+)$ as shown in Eq.~(\ref{fsru}).
Comparing the result in Eq.~(\ref{fb4})
with the $k_T$ resummation in Eq.~(\ref{fb}), which gives suppression, and
with the threshold resummation in Eq.~(\ref{fbt}), which gives 
enhancement, the unified resummation exhibits both behaviors: it is 
suppression in the large $b$ region ($p^+b\gg N$), and turns into 
enhancement in the small $b$ region ($N\gg p^+b$). That is, Eq.~(\ref{fb4})
displays the opposite effects of the $k_T$ and threshold resummations at
different $b$. The behavior of the unified resummation can be understood as
follows. For an intermediate $x$, virtual and real soft gluon corrections
cancel exactly in the small $b$ region, since they have almost equal phase
space. Hence, there are only single collinear logarithms, namely, no double
logarithms. In this case the Sudakov exponential approaches unity as
$b< 1/p^+$ \cite{LS}, indicating the soft cancellation stated above.
However, at threshold ($x\to 1$), real gluon emissions still do not have
sufficient phase space even as $b\to 0$, and soft virtual corrections are
not cancelled exactly. In this case the double logarithms $\ln^2(1/N)$
persist and become dominant. Sudakov suppression then transits into
enhancement, instead of unity, as $b$ decreases.

The unified resummation obtained in this work has important applications to
studies of some QCD processes, such as dijet production \cite{JH}. In this
experiment one jet (the trigger jet) is required to be in the central
rapidity region, while the other jet (the probe jet) has any rapidity up to
3.0. With the large rapidity, dynamics of a hadron at higher $x$ values is 
probed, so that the threshold resummation is necessary. On the other hand, 
the differential dijet cross section versus the transverse energy of the 
trigger jet is measured, which further demands the inclusion of the $k_T$
resummation. This subject will be discussed elsewhere.

\vskip 0.5cm
This work was supported by the National Science Council of Republic of
China under Grant No. NSC-88-2112-M-006-013.

\vskip 1.0cm

\centerline{\large\bf APPENDIX}
\vskip 0.5cm

In this Appendix
we present the details of the loop calculations involved in this paper.

(1) Equation (\ref{ts1})

The virtual gluon contribution to the soft function $K$ associated
with the threshold resummation for the quark distribution function is
given by
\begin{eqnarray}
K_s=ig^2C_F\mu^\epsilon\int\frac{d^{4-\epsilon}l}{(2\pi)^{4-\epsilon}}
N_{\nu\beta}(l)\frac{{\hat v}^\beta v^\nu}{v\cdot l}\frac{1}{l^2-a^2}
-\delta K\;,
\label{ass}
\end{eqnarray}
where the infrared regulator $a$ is introduced for convenience, and will
approach zero at last. The first term $g_{\nu\beta}$ in $N_{\nu\beta}$ gives
a vanishing contribution because of $v^2=0$. The contribution from the
second term $-n_\nu l_\beta/n\cdot l$ cancels that from the fourth term
$n^2l_\nu l_\beta/(n\cdot l)^2$. Hence, we concentrate only on the third
term $-n_\beta l_\nu/n\cdot l$, which leads to
\begin{equation}
K_s=-ig^2C_F\mu^\epsilon\int\frac{d^{4-\epsilon}l}{(2\pi)^{4-\epsilon}}
\frac{n^2}{(n\cdot l)^2}\frac{1}{l^2-a^2}-\delta K\;.
\label{ass1}
\end{equation}
The above integral has been evaluated in \cite{L1}, and thus we quote the
result directly:
\begin{equation}
K_s=\frac{\alpha_s(\mu)}{\pi}C_F\ln \frac{a}{\mu}\;,
\label{tsv1}
\end{equation}
with the counterterm $\delta K=-\alpha_sC_F/(\pi\epsilon)$.

The soft real gluon emission contributes
\begin{eqnarray}
K_r&=&ig^2C_F\int\frac{d^4l}{(2\pi)^4}
N_{\nu\beta}(l)\frac{{\hat v}^\beta v^\nu}{v\cdot l}
2\pi i\delta(l^2-a^2)\delta(1-z-l^+/p^+)\;,
\nonumber\\
&=&g^2C_F\int\frac{d^4l}{(2\pi)^3}
\frac{n^2}{(n\cdot l)^2}
2\pi \delta(l^2-a^2)\delta(1-z-l^+/p^+)\;,
\label{ass2}
\end{eqnarray}
with $z=x/\xi$, where $\epsilon$ has been set to zero, since the integral
is ultraviolet finite. Performing the integrations over $l^-$ and $l_T$,
Eq.~(\ref{ass2}) reduces to
\begin{eqnarray}
K_r=\frac{\alpha_s(\mu)}{\pi}C_F\frac{1-z}
{(1-z)^2-a^2/(2p^+\nu)^2}\;,
\label{tsr1}
\end{eqnarray}
with $\nu^2=(v\cdot n)^2/|n^2|$. Combined with Eq.~(\ref{tsv1}), the
function $K$ in moment space is, according to Eq.~(\ref{kmt}),
written as
\begin{eqnarray}
K(p^+/(N\mu),p^+)=\frac{\alpha_s(\mu)}{\pi}C_F\left[
\int_0^1 dz\frac{z^{N-1}(1-z)}
{(1-z)^2-a^2/(p^+\nu)^2}+\ln \frac{a}{\mu}\right]\;.
\label{tsr2}
\end{eqnarray}
A simple identity $z^{N-1}=(z^{N-1}-1)+1$ isolates the logarithmic term
$\ln a$ in the $z$ integral, and the above expression becomes
Eq.~(\ref{ts1}). As expected, the infrared regulator $a$ has
cancelled between the virtual and real gluon contributions.

(2) Equation (\ref{js1})

We need to evaluate only the real gluon part,
\begin{eqnarray}
K_r=ig^2C_F\int\frac{d^4l}{(2\pi)^4}N_{\nu\beta}(l)
\frac{{\hat v}^{\prime\beta} v^\nu}{v\cdot l}2\pi i\delta(l^2-a^2)
\delta(z-l^+/p^{\prime-}-wl^-/p^{\prime-})\;,
\label{b2}
\end{eqnarray}
with $z=(w-y)/(1-y)$. The first term $g_{\nu\beta}$ in $N_{\nu\beta}$ gives
a negligible contribution because of $v'^2=2w\to 0$. The contribution from
the second term $-n_\nu l_\beta/n\cdot l$ cancels that from the fourth term
$n^2l_\nu l_\beta/(n\cdot l)^2$. Hence, we consider the third
term $-n_\beta l_\nu/n\cdot l$, which leads to
\begin{equation}
K_r=g^2C_F\int\frac{d^4l}{(2\pi)^3}\frac{n^2}{(n\cdot l)^2}
\delta(l^2-a^2)\delta(z-l^+/p^{\prime-}-wl^-/p^{\prime-})\;.
\label{b3}
\end{equation}
Performing the integration over $l_T$ first, we obtain
\begin{equation}
K_r=g^2C_F\int\frac{dl^+dl^-}{8 \pi^2}
\frac{n^2}{(n\cdot l)^2}
\delta(z-l^+/p^{\prime-}-wl^-/p^{\prime-})\;,
\label{b4}
\end{equation}
with the constraint $2l^+l^-=l_T^2+a^2\ge a^2$. Here we have assumed
that the gauge vector $n$ possesses only the plus and minus components, 
{\it i.e.}, $n=(n^+,n^-.{\bf 0})$ for convenience. Using the relation
$l^+=zp^{\prime-}-wl^-$ from the $\delta$-function, the constriant
determines the bounds of $l^-$ in the integral
\begin{equation}
K_r=\frac{\alpha_s(\mu)}{2\pi}C_F\int_{l^-_{\rm min}}^{l^-_{\rm max}} dl^-
\frac{n^2p^{\prime-}}{[n^+l^-+n^-(zp^{\prime-}-wl^-)]^2}\;,
\label{b5}
\end{equation}
with
\begin{eqnarray}
& &l^-_{\rm min}=\frac{zp^{\prime-}-\sqrt{z^2p^{\prime-2}-2wa^2}}{2w}\;,
\nonumber\\
& &l^-_{\rm max}=\frac{zp^{\prime-}+\sqrt{z^2p^{\prime-2}-2wa^2}}{2w}\;.
\end{eqnarray}

The $l^-$ integration gives
\begin{eqnarray}
K_r&=&\frac{\alpha_s(\mu)}{\pi}C_F\frac{wn^2}{n^+-wn^-}
\nonumber \\
& &\times \Bigg\{\frac{1}{(n^++wn^-)z-(n^+-wn^-)
\sqrt{z^2-2wa^2/p^{\prime-2}}}
\nonumber \\
& &-\frac{1}{(n^++wn^-)z+(n^+-wn^-)
\sqrt{z^2-2wa^2/p^{\prime-2}}}\Bigg\}\;.
\end{eqnarray}
After a simple algebra, the above expression is simplified into
\begin{eqnarray}
K_r=\frac{\alpha_s(\mu)}{\pi}C_F\frac{\sqrt{z^2-2wa^2/p^{\prime-2}}}
{z^2+(n^+-wn^-)^2a^2/(n^2p^{\prime-2})}\;.
\label{b6}
\end{eqnarray}
Since the infrared regulator $a$ will approach zero at last, we drop
$a$ in the numerator, which does not affect the infrared structure of
$K_r$. According to the definition
\begin{equation}
\frac{(n^+-wn^-)^2}{n^2}=\frac{(n\cdot v')^2}{n^2}-v'^2\approx
\frac{(n\cdot v')^2}{n^2}\equiv-\nu'^2
\end{equation}
because of $n^2< 0$, Eq.~(\ref{b6}) becomes
\begin{eqnarray}
K_r=\frac{\alpha_s(\mu)}{\pi}C_F
\frac{z}{z^2-\nu'^2a^2/p^{\prime-2}}\;.
\label{b7}
\end{eqnarray}
Note that we should neglect the variable $w$ in $\nu'$ in order to decouple
the integrations over $y$ and over $w$ completely in Eq.~(\ref{jme}).

Combined with the virtual gluon contribution in Eq.~(\ref{tsv1}), we 
derive $K$ in moment space,
\begin{eqnarray}
K_r(p^{\prime-}/(N\mu),p^{\prime-})=\frac{\alpha_s(\mu)}{\pi}C_F
\left[\int_0^1 dz\frac{z^{N-1}(1-z)}{(1-z)^2-\nu'^2a^2/p^{\prime-2}}
+\ln\frac{a}{\mu}\right]\;,
\label{mj2}
\end{eqnarray}
where the variable change $z\to 1-z$ has been inserted. A similar
manipulation isolates $\ln a$ in the $z$ integral, and the
above expression leads to Eq.~(\ref{js1}).

(3) Equation (\ref{gj})

We evaluate first the soft subtraction term in Eq.~(\ref{gtj}),
\begin{eqnarray}
G_s&=&ig^2C_F\mu^\epsilon\int\frac{d^{4-\epsilon}l}{(2\pi)^{4-\epsilon}}
N_{\nu\beta}(l)\frac{{\hat v}^{\prime\beta}}{l^2}
\frac{2p^{\prime\nu}}{p'^2-2p'\cdot l}-\delta G\;,
\nonumber\\
&=&-ig^2C_F\mu^\epsilon\frac{n^2}{v'\cdot n}
\int\frac{d^{4-\epsilon}l}{(2\pi)^{4-\epsilon}}
\frac{2v'\cdot lp'\cdot n}{(n\cdot l)^2l^2(p'^2-2p'\cdot l)}-\delta G\;.
\label{gtjs}
\end{eqnarray}
We assume that the coefficient $n^+$ of $l^-$ in
the denominator $(n\cdot l)^2=(n^+l^-+n^-l^+)^2$ is negative. Since
the coefficient $-2wp'^-$ of $l^-$ in the denominator
$p'^2-2p'\cdot l=p'^2-2wp'^-l^--2p'^-l^+$ is also negative, a pole
from the denominator $l^2=2l^+l^--l_T^2$ exists in the $l^-$ plane
only for $l^+>0$. Performing the contour integration over $l^-$ around
the pole $l^-=l_T^2/(2l^+)$, Eq.~(\ref{gtjs}) gives
\begin{eqnarray}
G_s=2g^2C_Fn^2\mu^\epsilon
\int\frac{d^{2-\epsilon}l_T}{(2\pi)^{3-\epsilon}}
\int_0^\infty dl^+
\frac{l^+(2l^{+2}+wl_T^2)}{(2n^-l^{+2}+n^+l_T^2)^2
(2l^{+2}-2wp'^-l^++wl_T^2)}-\delta G\;.
\label{gtjs1}
\end{eqnarray}
The integration over $l^+$ leads to
\begin{eqnarray}
G_s&=&g^2C_F\mu^\epsilon
\int\frac{d^{2-\epsilon}l_T}{(2\pi)^{3-\epsilon}}
\frac{(n^+-wn^-)^2}{l_T^2(n^+-wn^-)^2+n^2w^2p^{\prime-2}}-\delta G\;.
\label{gtjs2}
\end{eqnarray}
At last, performing the integration over $l_T$, we derive 
\begin{eqnarray}
G_s=-\frac{\alpha_s(\mu)}{\pi}C_F
\ln\frac{wp^{\prime-}}{\nu'\mu}\;,
\label{gtjs3}
\end{eqnarray}
with the counterterm $\delta G=\alpha_sC_F/(\pi\epsilon)$.

Next we evaluate the first term in Eq.~(\ref{gtj}),
\begin{eqnarray}
G_h=ig^2C_F\frac{n^2}{v'\cdot n}\int\frac{d^4l}{(2\pi)^4}
\frac{(v'\cdot l)(\not p'-\not l)}{(n\cdot l)^2l^2(p'- l)^2}\not n\;.
\label{gtj1}
\end{eqnarray}
Note that we have set $\epsilon$ to zero, because the integral is
ultraviolet finite. It is easy to observe that poles of
$l^+$ exist in the denominator $l^2$ for $l^-<0$ and in
$(p'-l)^2$ for $l^-<p^{\prime-}$. Following the similar procedures,
we can work out Eq.~(\ref{gtj1}), though the calculation is much more
complicated. The final expression of $G_h$ is
\begin{eqnarray}
G_h=\frac{\alpha_s(\mu)}{2\pi}C_F\ln w\;,
\label{gtj2}
\end{eqnarray}
where constants of order unity have been dropped.
Combining Eqs.~(\ref{gtjs3}) and (\ref{gtj2}), we arrive at Eq.~(\ref{gj}).

A point that needs to be mentioned is the treatment of the gamma matrices
appearing in the numerator of Eq.~(\ref{gtj1}). We express $\gamma^+$ and
$\gamma^-$ in terms of $\not p'$ and $\not n$:
\begin{eqnarray}
\gamma^+=\frac{n^+\not p'-wp^{\prime-}\not n}{p^{\prime-}(n^+-wn^-)}\;,
\;\;\;\;
\gamma^-=\frac{p^{\prime-}\not n-n^-\not p'}{p^{\prime-}(n^+-wn^-)}\;.
\end{eqnarray}
Since the jet function contains the matrix structure proportional
to $\not p'$, the products $\gamma^+\not n$ and $\gamma^-\not n$ are
written as
\begin{eqnarray}
\gamma^+\not n=\frac{2n^+p'\cdot n-wp^{\prime-}n^2}{p^{\prime-}(n^+-wn^-)}\;,
\;\;\;\;
\gamma^-\not n=\frac{p^{\prime-}n^2-2n^-p'\cdot n}{p^{\prime-}(n^+-wn^-)}\;,
\label{npp}
\end{eqnarray}
where we have used the relation $\not p'\not n=2p'\cdot n-\not n \not p'$,
and neglected the second term, which leads to a negligible invariant
$p^{\prime 2}=wp^{\prime-2}\approx 0$.


\newpage
\centerline{\large \bf Figure Captions}
\vskip 0.3cm

\noindent
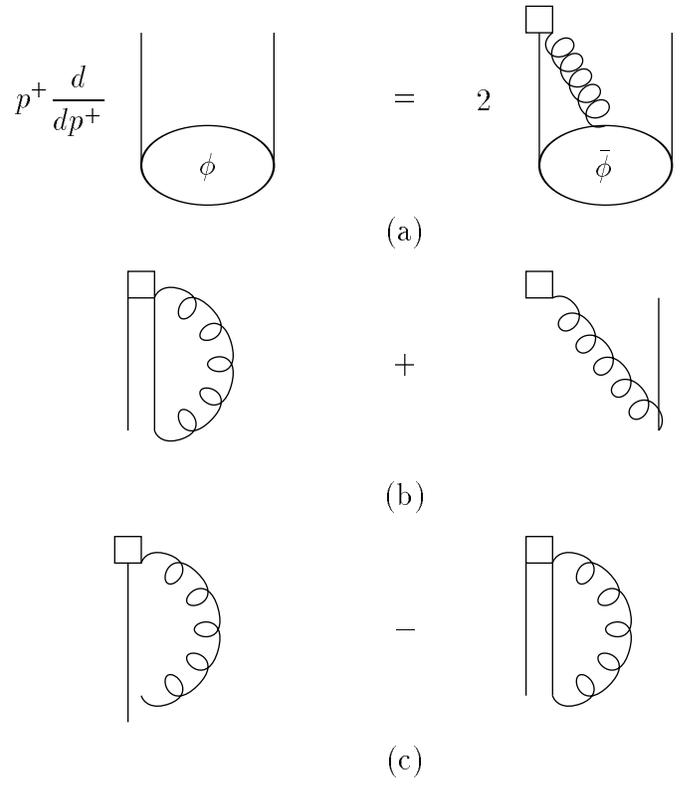
{\bf Fig. 1.} (a) The derivative $p^+d\phi/dp^+$ in the axial gauge. (b)
The $O(\alpha_s)$ function $K$. (c) The $O(\alpha_s)$ function $G$.
\vskip 0.3cm

\end{document}